\documentclass[12pt]{article}
\usepackage{amssymb}

\input{tcilatex}

\begin{document}

\begin{center}
\textbf{PROPAGATION AND INTERACTION OF EXTREMELY SHORT ELECTROMAGNETIC
PULSES IN NON-LINEAR MEDIA}

\vspace{0.5cm} A.I.Maimistov

\vspace{0.2cm}Moscow Engineering Physics Institute, Kashirskoe sh. 31,
Moscow, Russia,

e-mail: maimistov@pico.mephi.ru \bigskip

\textbf{ABSTRACT}
\end{center}

\bigskip

Propagation of the extremely short electromagnetic pulse in non-linear
dielectric media without the slowly varying envelope approximation is
discussed. The models under consideration take into account both resonant
and not-resonant excitations of non-linear medium, and polarisation states
of electromagnetic wave. Steady state solutions of the relevant equations
are presented for certain of these models.

\bigskip


\newpage

\section{Introduction}

In the context of rapid progress in the generation of femtosecond pulses of
the electromagnetic radiation, it is of interest to consider the theoretical
models of the pulses propagation in non-linear dispersive medium. Indeed,
derivation of the evaluation equations for electromagnetic radiation is
often based on the approximation of slowly varying complex envelopes of the
optical pulses. This approximation is employed widely in theoretical studies
devoted to non-linear coherent phenomena. However, it is of interest to find
a method of describing extremely short pulse (ESP) evolution \emph{without
using the slowly varying envelope approximation}.

A natural foundation for all the theories is the Maxwell equations
supplemented with equations governing the evolution of the polarisation or
currents arising in a medium subject to electromagnetic radiation. Since it
seldom occurs that exact analytical results can be obtained in such
problems, various approximations simplifying the consideration and allowing
analytical expressions to be derived are often employed.

The typical models of the non-linear medium are the ensembles of $N$-level
atoms or anharmonic oscillators. So, we can discuss the different degrees of
approximation in the theory of interaction and propagation of optical pulses
in framework of the resonant and non-resonant representation for medium \cite
{R1}.

\section{Extremely short pulse propagation in resonance medium}

Let the spectral width of ESP be far smaller than the resonance frequency.
This resonant condition allows to omit all non-resonant levels and to
consider only two levels. Thus, we obtain \emph{the approximation of the
two-level atoms} \cite{R2}, which is very popular in the resonant and
coherent optics. The wave equation in this case is complemented by the Bloch
equations for two-level atom variables.

\subsection{Propagation of the scalar ESP}

In general, the theory of the interaction of radiation with an ensemble of
two-level atoms is based on the Bloch equations for atoms and the Maxwell
equations for the classical electromagnetic field. In an isotropic
dielectric the set of Maxwell equations reduced to one equation for the
electric field $\vec{E}=E\vec{l}$. For a plane wave with constant
polarisation vector one can obtain the following system of total
Maxwell-Bloch (MB) equations

\begin{equation}
\frac{\partial ^{2}E}{\partial z^{2}}-\frac{1}{c^{2}}\frac{\partial ^{2}E}{%
\partial t^{2}}=\frac{4\pi n_{A}d}{c^{2}}\left\langle \frac{\partial
^{2}r_{1}}{\partial t^{2}}\right\rangle   \label{eq1a}
\end{equation}

\begin{equation}
\frac{\partial r_{1}}{\partial t}=-\omega _{a}r_{2},\quad \frac{\partial
r_{2}}{\partial t}=\omega _{a}r_{1}+\frac{2d}{\hbar }Er_{3},\quad \frac{%
\partial r_{3}}{\partial t}=-\frac{2d}{\hbar }Er_{2},  \label{eq1b}
\end{equation}
where $d$\ is the projection of a matrix element of the dipole operator on
the direction of $\vec{l}$, $n_{A}$ is the concentration of resonant atoms.
It should be noted that the components of Bloch vector $r_{1}$, $r_{2}$ ,
and $r_{3}$ depend on the atomic resonance frequency $\omega _{a}$.
Hereafter the angular brackets represent summation over all the atoms
characterised by the frequency $\omega _{a}$.

Let us consider the solution of equations (\ref{eq1a}) and (\ref{eq1b})
describing a solitary steady-state wave. Boundary conditions at $%
|t|\rightarrow \infty $ must look like the following 
\[
E=\partial E/\partial t=0,\quad r_{2}=r_{1}=0,\quad r_{3}=-1. 
\]

To obtain the expressions describing the propagation of stationary ESP one
should suppose that the components of the Bloch vector and the electric
field $E$ depend on only one variable $t\pm z/V$ or $\eta =\omega _{a}(t\pm
z/V)$ \ in dimensionless form. That means that a stationary (steady state)
wave propagates only in one (positive or negative) direction. Under this
assumption the system (\ref{eq1a}), (\ref{eq1b}) transforms into the system
of ordinary differential equations which can be solved by usual methods.
Solution of the these ordinary differential equations leads to the
expression for electric field strength 
\begin{equation}
E(z,t)=E_{0}\sec {\normalsize h}\left[ dE_{0}(t\pm z/V)/\hbar \right] ,
\label{eq2}
\end{equation}
where pulse amplitude is related with pulse duration $t_{p}$ as $E_{0}=\hbar
t_{p}^{-1}d^{-1}$. This expression for stationary pulse correlates with the
one found in \cite{R3, R4}. Expression for the stationary ESP velocity can
be obtained from definitions of pulse duration and amplitude. As in
self-induced theory by McCall-Hahn \cite{R2, R5}, these values are related
to each other, so the pulse is able to invert a two-level system and then
return it to the initial state while the pulse lasts. After some algebra one
may find 
\begin{equation}
\frac{1}{V^{2}}=\frac{1}{c^{2}}\left( 1+\frac{8\pi n_{A}d^{2}\hbar \omega
_{a}}{(dE_{0})^{2}+(\hbar \omega _{a})^{2}}\right) ,  \label{eq3}
\end{equation}
that corresponds to what was found in \cite{R3}.

\subsection{Propagation of the ESP of polarised radiation}

In this case generalised Maxwell-Bloch equations can be written as follows 
\begin{equation}
\frac{\partial ^{2}E^{(+1)}}{\partial z^{2}}-\frac{1}{c^{2}}\frac{\partial
^{2}E^{(+1)}}{\partial t^{2}}=\frac{4\pi n_{A}}{c^{2}}\frac{\partial ^{2}}{%
\partial t^{2}}\left\langle d_{13}\rho _{31}+d_{31}\rho _{13}\right\rangle ,
\label{eq4a}
\end{equation}
\begin{equation}
\frac{\partial ^{2}E^{(-1)}}{\partial z^{2}}-\frac{1}{c^{2}}\frac{\partial
^{2}E^{(-1)}}{\partial t^{2}}=\frac{4\pi n_{A}}{c^{2}}\frac{\partial ^{2}}{%
\partial t^{2}}\left\langle d_{23}\rho _{32}+d_{32}\rho _{23}\right\rangle ,
\label{eq4b}
\end{equation}
\[
i\hbar \frac{\partial \rho _{13}}{\partial t}=-\hbar \omega _{a}\rho
_{13}+d_{13}(\rho _{33}-\rho _{11})E^{(+1)}-d_{23}\rho _{12}E^{(-1)}, 
\]
\[
i\hbar \frac{\partial \rho _{23}}{\partial t}=-\hbar \omega _{a}\rho
_{23}+d_{23}(\rho _{33}-\rho _{22})E^{(-1)}-d_{13}\rho _{21}E^{(+1)}, 
\]
\begin{equation}
i\hbar \frac{\partial \rho _{12}}{\partial t}=d_{13}\rho
_{32}E^{(+1)}-d_{32}\rho _{13}E^{(-1)},  \label{eq4c}
\end{equation}
\[
i\hbar \frac{\partial }{\partial t}(\rho _{11}-\rho _{33})=2(d_{13}\rho
_{31}-d_{31}\rho _{13})E^{(+1)}-(d_{23}\rho _{32}-d_{32}\rho _{23})E^{(-1)}, 
\]
\[
i\hbar \frac{\partial }{\partial t}(\rho _{22}-\rho _{33})=(d_{13}\rho
_{31}-d_{31}\rho _{13})E^{(+1)}-2(d_{23}\rho _{32}-d_{32}\rho
_{23})E^{(-1)}, 
\]
where $E^{(q)}$ is the spherical $q$ component of the vector of electrical
field strength of a light wave, $q=\pm 1$, $d_{kl}$\ are the matrix elements
of the dipole moment operator of the atomic transition. Hereafter $%
d_{13}=d_{23}=d_{31}^{\ast }=d_{32}^{\ast }$.

By supposing that the matrix elements of the density matrix $\rho _{kl}$,
and the electric field $E$\ depend on only one variable $t\pm z/V$ we can
find the expression for the strength of the electrical field of the ESP in
the form 
\begin{equation}
E^{(\pm 1)}(z,t)=e^{(\pm )}E_{0}\sec {\normalsize h}\left[ dE_{0}(t\pm
z/V)/\hbar \right]  \label{eq5}
\end{equation}
where $e^{(\pm )}$ are the components of a unit vector of electromagnetic
fields. This is a simple generalisation of the results obtained by Bullough
and Ahmad \cite{R4} to the case of vector (polarised) ESP and specific model
of a resonant medium. As in Ref. \cite{R4}, the pulse duration tp is
expressed in terms of the pulse peak amplitude $t_{p}=\hbar (dE_{0})^{-1}$.
The propagation velocity of the ESP is: 
\begin{equation}
\frac{1}{V^{2}}=\frac{1}{c^{2}}\left( 1+\frac{8\pi n_{A}d^{2}\hbar \omega
_{a}}{2(dE_{0})^{2}+(\hbar \omega _{a})^{2}}\right) .  \label{eq6}
\end{equation}

Besides the solitary wave solutions, equations (\ref{eq1a}), (\ref{eq1b})
and (\ref{eq4a})-(\ref{eq4c}) have the solutions in the form of cnoidal
wave. 

The above analysis of the ESP propagation shows that the nature of such
stationary ESP depends on the state of the medium. The scalar solution \cite
{R4} can be trivially generalised to cover a vector video pulse case under
condition $n_{10}=n_{20}=1$. So we obtain a circular polarised pulse \cite
{R6} with duration of a half of a reciprocal atomic transition frequency.
The propagation velocity of this pulse does not depend on the conditions of
its polarisation and the value of this velocity is in agreement with what
was found in \cite{R4}.

The new solution of the generalised complete Maxwell-Bloch equations arises
when initially the resonance medium has an asymmetrical distribution of
level populations (i.e., $n_{10}$ and $n_{20}$) of the excited states that
have different projections of the angular momentum: $n_{10}\neq n_{20}$ \cite
{R7}. One of the spherical components of the electric field strength vector
behaves as in the scalar case, i.e., it is a unipolar spike of the
electrical field. Another component is a variable-sign solitary wave. All
stationary solutions like these form a three-parametric family, where each
member is defined by the propagation speed $V$ and $n_{10}$, $n_{20}$
determining the polarisation state of the pulse.

\section{Extremely short pulse propagation under quasi-resonance condition}

Let assume that the amplitude of electrical field of electromagnetic pulse
is so small that the instant Rabi frequency $\omega _{R}$\ proves to be far
less then the frequency of the resonance transition, i.e., 
\begin{equation}
\varepsilon =\frac{\omega _{R}}{\omega _{a}}=\frac{|d_{13}|E_{0}}{\hbar
\omega _{a}}\ll 1,  \label{eq7}
\end{equation}
where $\omega _{R}$ corresponds to the wave of constant electrical field
amplitude $E_{0}=\max |E^{(q)}|$. In this case the solution of Bloch
equations ( \ref{eq4c}) can be presented formally \cite{R8} by power series
in $\varepsilon $. Limiting ourselves to the terms of this series up to $%
\varepsilon ^{3}$, for polarisation in equations (\ref{eq4a}) and (\ref{eq4b}%
), $P^{(+1)}=$\ $d_{13}\rho _{31}+d_{31}\rho _{13},$\ $P^{(-1)}=$\ $%
d_{23}\rho _{32}+d_{32}\rho _{23}$,\ \ we obtain expressions \cite{R8} : 
\begin{equation}
P^{(q)}=-\frac{2\sigma |d_{13}|^{2}}{\hbar \omega _{a}}\left\{ E^{(q)}-\frac{%
1}{\omega _{a}^{2}}\frac{\partial ^{2}E^{(q)}}{\partial t^{2}}-\frac{%
2|d_{13}|^{2}}{(\hbar \omega _{a})^{2}}(\vec{E}\cdot \vec{E})E^{(q)}\right\}
.  \label{eq8}
\end{equation}
The substitution of (\ref{eq8}) into equations (\ref{eq4a}) and (\ref{eq4b})
results in the non-linear wave equation, which gives approximate description
of the extremely-short pulse of electromagnetic wave 
\begin{equation}
\frac{\partial ^{2}\vec{E}}{\partial z^{2}}-\frac{1}{V^{2}}\frac{\partial
^{2}\vec{E}}{\partial t^{2}}-b\frac{\partial ^{4}\vec{E}}{\partial t^{4}}-a%
\frac{\partial ^{2}}{\partial t^{2}}\{(\vec{E}\cdot \vec{E})\vec{E}\}=0
\label{eq9}
\end{equation}
where 
\[
a=\left\langle \frac{16\pi n_{A}\sigma |d_{13}|^{2}}{c^{2}(\hbar \omega
_{a})^{3}}\right\rangle ,\quad b=\left\langle \frac{8\pi n_{A}\sigma
|d_{13}|^{2}}{c^{2}\hbar \omega _{a}{}^{3}}\right\rangle . 
\]
The velocity of pulse propagation $V$ alters because of the dispersion,
which is due to the resonant medium: 
\[
V^{-2}=c^{-2}\left[ 1-\left\langle 8\pi n_{A}\sigma |d_{13}|^{2}/\hbar
\omega _{a}\right\rangle \right] . 
\]

For zero boundary conditions as \ $t\rightarrow \pm \infty $ the steady
state solution of equation (\ref{eq9}) can be written in the form: 
\begin{equation}
\vec{E}(z,t)=\vec{e}E_{0}\sec \mathrm{h}\left[ (t\pm \alpha z/V-t_{0})/t_{S}%
\right]  \label{eq10}
\end{equation}
where 
\[
t_{S}=E_{0}^{-1}\sqrt{|b|/2|a|},\quad \alpha =\sqrt{1+2\sigma
E_{0}^{2}V^{2}|a|}. 
\]

Thus, the steady state solutions of the equation (\ref{eq9}) describe the
extremely short pulse propagation with constant polarisation state (i.e., $%
\vec{e}=const$ ) of this electromagnetic wave.

Spectral half-width of the ESP (\ref{eq10}) is $t_{S}^{-1}$. Consequently
the resonance condition will be applicable if $t_{S}^{-1}\ll \bar{\omega}$.
From the definition of the parameters \TEXTsymbol{\vert}$a|$ and $|b|$ we
can estimate their ratio as $|a|/|b|\approx 2|d_{13}|^{2}\bar{\omega}%
^{2}\hbar ^{-2}$\ (where we neglected inhomogeneous broadening and all the
frequencies $\omega _{a}$\ were replaced by $\bar{\omega}$). Thus, the
resonance condition complies with the condition (\ref{eq7}) $\omega _{R}\ll 
\bar{\omega}$.

Besides the solitary steady state wave (\ref{eq10}) both dark solitary waves
and a periodic stationary waves can be found. However they contradict with
conditions of applicability of the initial equations for atomic variables,
in where any relaxation processes are not taking into account. The
considering pulse is short one in this sense.

\section{Extremely short pulse propagation in non-resonance medium}

As it was pointed in \cite{R9} when the ESP field approaches the atomic
field ($\thicksim 10^{8}$ - $10^{9}$ V/cm), ESP formation is affected mainly
by the atomic ionisation potential, which limits pulse duration and
amplitude. For such fields the resonant models are invalid. However, one can
evaluate limitations on ESP within a classical model of an atom with a
strong non-linear potential, limited at infinity, to allow for ionisation.
Then the (generalised) Bloch equations are replaced by classical equations
for the electron (and ion) motion.

\subsection{Scalar Duffing model}

Let the medium be a non-linear and non-resonant one. A similar approach was
exploited \cite{R10} to develop the self-induced transparency in ionic
crystals in the framework of the Duffing's type model. The scalar wave
equation is 
\begin{equation}
\frac{\partial ^{2}E}{\partial z^{2}}-\frac{1}{c^{2}}\frac{\partial ^{2}E}{%
\partial t^{2}}=\frac{4\pi }{c^{2}}\frac{\partial ^{2}P}{\partial t^{2}}.
\label{eq11}
\end{equation}
We complete this wave equation by the non-linear oscillator equation for the
medium. If $X$ is the displacement of an electron from its equilibrium
position, the motion equation (which neglects friction) can be written as 
\begin{equation}
\frac{\partial ^{2}X}{\partial t^{2}}+\omega _{0}^{2}X+\kappa _{3}X^{3}=%
\frak{L}\frac{e}{m}E  \label{eq12}
\end{equation}
where $\omega _{0}$\ is an eigenfrequency of the oscillator, while $\kappa
_{3}$ is anharmonicity coefficient. The term on the right-hand side of
equation (\ref{eq12}) represents the force exerted on the electron by the
electromagnetic field, where $\frak{L}=(\varepsilon +2)/3$ is the Lorentz
factor, $e$ is the electron's electric charge, and $m$ is its mass. However,
it is possible to absorb $\frak{L}$\ into an effective mass $m_{eff}=m/\frak{%
L}$. Hereafter, we will use $m$ as a symbol for this effective mass.
Finally, the dynamical variable $X$ is related to the medium's polarisation, 
$P=n_{A}eX$, where $n_{A}$ is the density of the oscillators (atoms).

The steady state solutions of the system of equations (\ref{eq11}) and (\ref
{eq12}) can be found if we suppose that the displacement $X$ and electric
strength $E$ depend on one variable $\eta =\omega _{0}(t\pm z/V)$. Here
velocity $V$ is a parameter of the solution. Thus, the expression for
electric field of ESP can be obtained by solving the ordinary non-linear
differential equations. That leads to 
\begin{equation}
E(t\pm z/V)=E_{m}\alpha (\alpha -1)^{1/2}\sec \mathrm{h}\left[ (\alpha
-1)^{1/2}\omega _{0}(t\pm z/V-t_{0})\right]  \label{eq13}
\end{equation}
where $E_{m}=(m\omega _{0}^{3}/e)\sqrt{2/\kappa _{3}}$, $\alpha =(\omega
_{p}/\omega _{0})^{2}V^{2}/(c^{2}-V^{2})$, and $\omega _{p}=(4\pi
n_{A}e^{2}/m)^{1/2}$\ \ is the plasma frequency:

\subsection{Vector Duffing model}

Let us consider the ESP propagation in a non-linear medium, which is again
presented by an ensemble of the molecules. However, now let us suppose that
the inner degrees of freedom are described by the potential: 
\begin{equation}
U(X,Y)=\frac{1}{2}\omega _{1}X^{2}+\frac{1}{2}\omega _{2}Y^{2}+\frac{1}{2}%
\kappa _{2}X^{2}Y^{2}+\frac{1}{4}\kappa _{4}(X^{4}+Y^{4}),  \label{eq14}
\end{equation}
where $\kappa _{2}$\ and $\kappa _{4}$\ are anharmonicity coefficients. If
the coupling parameter $\kappa _{2}=0$, then this potential describes the
scalar Duffing model of anharmonic oscillator, which can oscillate
independently in two orthogonal directions. Expression (\ref{eq14}) is the
simplest generalisation of this model. Polarisation of the molecule is
defined by the expression $\vec{p}=eX\vec{e}_{1}+eY\vec{e}_{2}$, and the
total polarisation $\vec{P}$ is the product of the density $n_{A}$ and the
polarisation of one molecule.

The propagation of the ESP will be considered in the unidirectional wave
approximation, so the wave equations for electric field corresponding to the
different polarisation components of the ESP can be written as follows 
\begin{equation}
\frac{\partial ^{2}E_{1}}{\partial z^{2}}-\frac{1}{c^{2}}\frac{\partial
^{2}E_{1}}{\partial t^{2}}=\frac{4\pi n_{A}e}{c^{2}}\frac{\partial ^{2}X}{%
\partial t^{2}},\quad \frac{\partial ^{2}E_{2}}{\partial z^{2}}-\frac{1}{%
c^{2}}\frac{\partial ^{2}E_{2}}{\partial t^{2}}=\frac{4\pi n_{A}e}{c^{2}}%
\frac{\partial ^{2}Y}{\partial t^{2}}.  \label{eq15}
\end{equation}
The equations of motion for anharmonic oscillator follow from the classical
Newton equations 
\begin{equation}
\frac{\partial ^{2}X}{\partial t^{2}}+\omega _{1}^{2}X+\kappa
_{2}XY^{2}+\kappa _{4}X^{3}=\frac{e}{m}E_{1},  \label{eq16}
\end{equation}
\begin{equation}
\frac{\partial ^{2}Y}{\partial t^{2}}+\omega _{2}^{2}Y+\kappa
_{2}YX^{2}+\kappa _{4}Y^{3}=\frac{e}{m}E_{2}.  \label{eq17}
\end{equation}

It is possible to find a steady state solution of this system of equations.
Let us suppose the fields depend only on $\tau =t\pm z/V$. Integrating
equations (\ref{eq15}) and taking account of the boundary conditions, which
correspond to the vanishing electric field of the USP and polarisation of
molecules at infinity, we find that $E_{1}=4\pi n_{A}e\theta X$\ and $%
E_{2}=4\pi n_{A}e\theta Y$. Substitution of this result into (\ref{eq16})
and (\ref{eq17}) leads to the following equations 
\begin{equation}
\frac{d^{2}X}{d\tau ^{2}}+(\kappa _{2}Y^{2}+\kappa _{4}X^{2})X=(\theta
\omega _{p}^{2}-\omega _{1}^{2})X,  \label{eq18a}
\end{equation}
\begin{equation}
\frac{d^{2}Y}{d\tau ^{2}}+(\kappa _{2}X^{2}+\kappa _{4}Y^{2})Y=(\theta
\omega _{p}^{2}-\omega _{2}^{2})X,  \label{eq18b}
\end{equation}
where $\theta =V^{2}(c^{2}-V^{2})^{-1}$\ These equations appear in many
works, so we can easily find their solutions. Let suppose that the
eigenfrequencies $\omega _{1}$\ and $\omega _{2}$ are equal. The partial
solution of system (\ref{eq18a}), (\ref{eq18b}) in this case is 
\[
E_{1}(\tau )=E_{2}(\tau )=E_{m}\alpha (\alpha -1)^{1/2}\sec \mathrm{h}%
[(\alpha -1)^{1/2}\omega _{1}(\tau -\tau _{0})], 
\]
where $E_{m}=(m\omega _{1}^{3}/e)\sqrt{2/(\kappa _{2}+\kappa _{4})},$ $\tau
_{0}$\ is the integration constant. This solution similar to that above
describes the linear polarised video pulse propagating without distortion in
the considered non-linear medium.

The more interesting solution exists if the frequencies $\omega _{1}$\ and $%
\omega _{2}$ are different, but factors of anharmonicity coincide $\kappa
_{2}=\kappa _{4}$. In this case the solution of the system of equations (\ref
{eq18a}) and (\ref{eq18b}) leads to following expressions \cite{R11}: 
\[
E_{1}(\tau )=\frac{8\pi en_{A}\theta \sqrt{2\kappa _{2}^{-1}}\mu _{1}\exp
(\theta _{1})\{1+\exp (2\theta _{2}+\mu _{12})\}}{1+\exp (2\theta _{1})+\exp
(2\theta _{2})+\exp (2\theta _{1}+2\theta _{2}+\mu _{12})}, 
\]
\begin{equation}  \label{eq19}
\end{equation}
\[
E_{2}(\tau )=\frac{8\pi en_{A}\theta \sqrt{2\kappa _{2}^{-1}}\mu _{2}\exp
(\theta _{2})\{1+\exp (2\theta _{1}+\mu _{12})\}}{1+\exp (2\theta _{1})+\exp
(2\theta _{2})+\exp (2\theta _{1}+2\theta _{2}+\mu _{12})}, 
\]
where 
\[
\exp \mu _{12}=(\mu _{1}-\mu _{2})(\mu _{1}+\mu _{2})^{-1}. 
\]
In these expressions the following parameters $\mu _{1,2}=(\theta \omega
_{p}^{2}-\omega _{1,2}^{2})$\ and $\theta _{1,2}=\mu _{1,2}(\tau -\tau
_{1,2})$\ are used, where $\tau _{1,2}$ are the integration constants.
Solutions (\ref{eq19}) describe the steady state propagation of the ESP
where one of these components corresponds to a \emph{unipolar} spike of the
electric field, sub-cycle pulse. The other component is a \emph{bipolar}
impulse or monocycle pulse.

\subsection{Cubic-quadratic Duffing model}

An objective of the present section is to discuss the propagation and
interactions of linearly polarised USPs in a non-linear dispersive medium
modelled by an anharmonic oscillator combining quadratic and cubic
nonlinearities. As is well known, in the case when the oscillator is
quasi-harmonic, the quadratic and cubic non-linear terms produce effects of
the same order of magnitude \cite{R12}, that is why it is natural to
consider a mixed model of this type. It is also relevant to mention that
modelling dynamics of the broad (rather than ultra-short) optical solitons
in a medium with competing quadratic and cubic nonlinearities has recently
attracted considerable attention, see, e.g., the works \cite{R13} and
references therein.

The one-dimensional propagation of the electromagnetic waves in a non-linear
medium is governed by the wave equation (\ref{eq11}). We adopt a simple
anharmonic oscillator model for the medium, which is a frequently used
approximation \cite{R14} (see also \cite{R15, R16}). If $X$ is the
displacement of an electron from its equilibrium position, the motion
equation (which neglects friction) can be written as 
\begin{equation}
\frac{\partial ^{2}X}{\partial t^{2}}+\omega _{0}^{2}X-\kappa
_{2}X^{2}+\kappa _{3}X^{3}=\frac{e}{m}E,  \label{eq20}
\end{equation}
where $\kappa _{2}$\ and $\kappa _{4}$\ are anharmonicity coefficients. It
seems plausible that the system of equations (\ref{eq11}), (\ref{eq20}) is
not an integrable one. Nevertheless, some exact analytical solutions,
describing the propagation of ESPs, can be found. To this end, one assumes
that $E$ and $X$ depend on a single variable, $\eta =\omega _{0}(t\pm z/V)$\
with some velocity $V$. It is convenient to express the velocity $V$ of a
steadily moving pulse by introducing the constant $\alpha $ 
\begin{equation}
\alpha =\left( \frac{\omega _{p}}{\omega _{0}}\right) ^{2}\frac{V^{2}}{%
c^{2}-V^{2}},  \label{eq21}
\end{equation}
where $\omega _{p}$\ is the plasma frequency. The first equation of the
system, i.e., (\ref{eq11}), can be integrated. It allow us to obtain a
family of exact solutions parameterised by the continuous \emph{positive}
parameter $(\alpha -1)$ and discrete one $\sigma =\pm 1$, 
\begin{equation}
E^{(+)}(\eta ;\alpha )=\frac{3E_{m}\alpha (\alpha -1)}{\sqrt{1+9(\alpha
-1)\mu }\cosh \left( \sqrt{(\alpha -1)}\eta \right) -1},  \label{eq22}
\end{equation}
\begin{equation}
E^{(-)}(\eta ;\alpha )=-\frac{3E_{m}\alpha (\alpha -1)}{\sqrt{1+9(\alpha
-1)\mu }\cosh \left( \sqrt{(\alpha -1)}\eta \right) +1},  \label{eq23}
\end{equation}
where $E_{m}=(m\omega _{0}^{4}/e|\kappa _{2}|)$, $\mu =(\kappa _{3}\omega
_{0}^{2}/2\kappa _{2}^{2}),$\ and the superscript standing for $\sigma $.
The pulses represented by the solutions (\ref{eq27}) and (\ref{eq28}) have 
\emph{different polarities}. In the limit $\kappa _{3}\rightarrow 0$,
corresponding to the model with a purely quadratic nonlinearity, equation (%
\ref{eq22}) goes over into a singular solution, 
\[
E^{(+)}(\eta ;\alpha )=\frac{3E_{m}\alpha (\alpha -1)}{2\sinh ^{2}\left( 
\sqrt{(\alpha -1)}\eta /2\right) }, 
\]
while expression (\ref{eq23}) yields a non-singular pulse in the same limit,
which was already found in Ref. \cite{R17}, 
\[
E^{(-)}(\eta ;\alpha )=-\frac{3E_{m}\alpha (\alpha -1)}{2\cosh ^{2}\left( 
\sqrt{(\alpha -1)}\eta /2\right) }. 
\]
Thus, in compliance with the results of Ref. \cite{R17}, only one family of
non-singular pulses exists in the case of the purely quadratic nonlinearity.
We stress that the velocities of two pulses, which have the opposite
polarities but the same value of $\alpha $\ are equal, since they are
determined by equation (\ref{eq21}) and depend only on $\alpha $.

Although the pulse solutions have been obtained in the exact form, their
stability and interactions should be studied by means of numerical
simulations. This investigation was performed under unidirectional
approximation. Simulations of interactions between have shown that their
collisions are quasi-elastic, irrespective of the polarities of the
colliding pulses: after passing through each other, the pulses retrieve the
same shapes and velocities as they had before the collision, the only result
being a shift of their centers. However, the character of the interaction
between pulses with equal and opposite polarities becomes different with the
decrease of their relative velocity. In the case of equal polarities, the
pulses with a small relative velocity demonstrate strong mutual repulsion:
the distance between them attains a minimum, and then they start to separate
again, so that they never completely overlap. Strong energy exchange between
the pulses takes place around the point where they attain the minimum
separation. The energy exchange gives rise to mutual interconversion of the
two pulses, so that after the collision they, effectively, swap their
positions.

\subsection{Vector model of an quadratic non-linear media}

The propagation of an ESP of polarised radiation in a medium with a
quadratic nonlinearity can be considered by using a model of an ensemble of
two-component anharmonic oscillators \cite{R18}. This model described by the
wave equations (\ref{eq15}) supplemented with the following equations 
\begin{equation}
\frac{\partial ^{2}X}{\partial t^{2}}+\omega _{1}^{2}X+\kappa
_{111}X^{2}+2\kappa _{112}XY+\kappa _{122}Y^{2}=\frac{e}{m_{1}}E_{1},
\label{eq24a}
\end{equation}
\begin{equation}
\frac{\partial ^{2}Y}{\partial t^{2}}+\omega _{2}^{2}Y+\kappa
_{112}X^{2}+2\kappa _{122}XY+\kappa _{222}Y^{2}=\frac{e}{m_{2}}E_{2},
\label{eq24b}
\end{equation}
where the anharmonicity coefficients $\kappa _{ijl}$ and two effective
masses $m_{1}$\ and $m_{2}$ are introduced. Let the electric field strengths
and displacements $X$ and $Y$ depend only on the variable $\tau =t\pm z/V$.
Similar to the scalar case, one can integrate the wave equations (\ref{eq15}%
) and equations (\ref{eq24a}), (\ref{eq24b}) can be reduced into the
following set of non-linear equations 
\begin{equation}
\frac{d^{2}X}{d\tau ^{2}}+(\omega _{1}^{2}-\theta \omega _{p}^{2})X+\kappa
_{111}X^{2}+2\kappa _{112}XY+\kappa _{122}Y^{2}=0,  \label{eq25a}
\end{equation}
\begin{equation}
\frac{d^{2}Y}{d\tau ^{2}}+(\omega _{2}^{2}-\theta \omega _{p}^{2})Y+\kappa
_{112}X^{2}+2\kappa _{122}XY+\kappa _{222}Y^{2}=0.  \label{eq25b}
\end{equation}
If both the eigenfrequencies of oscillators and the anharmonicity
coefficients are equal to each other, then the problem considered can by
reduced to the scalar one. Unfortunately, the general solution of the
equations (\ref{eq25a}), (\ref{eq25b}) is unknown.

\subsection{Coupled between vibrations and light waves in Raman media}

To description of the light propagation in Raman medium in \cite{R19} the
oscillator model has been proposed. The displacement $Q$, vibration density $%
\rho $, and electric field of the light wave $E$ are governed by the
following equations 
\[
\frac{\partial ^{2}E}{\partial z^{2}}-\frac{1}{c^{2}}\frac{\partial ^{2}E}{%
\partial t^{2}}=\frac{4\pi }{c^{2}}\frac{\partial ^{2}P}{\partial t^{2}}, 
\]
\begin{equation}
\frac{\partial ^{2}Q}{\partial t^{2}}+\frac{1}{T_{2}}\frac{\partial Q}{%
\partial t}+\omega _{0}^{2}Q=-\frac{\alpha _{1}}{2M}E^{2}\rho ,  \label{eq26}
\end{equation}

\[
\frac{\partial \rho }{\partial t}+\frac{1}{T_{1}}(\rho -\rho _{0})=-\frac{%
\alpha _{1}}{\hbar \omega _{0}}E^{2}\frac{\partial Q}{\partial t}, 
\]
where $\omega _{0}$ is an eigenfrequency of the oscillator,$M$ is a mass of
ion. The non-linear polarisation is defined by expression $P=n_{A}\alpha
_{1}QE$, where parameter $\alpha _{1}$\ is the coefficient in expansion of
the atomic (or molecular) polarizability $\alpha (Q)\approx \alpha
_{0}+\alpha _{1}Q$. The relaxation processes were taken into account by
introduction relaxation times $T_{1}$ and $T_{2}$. The constraint $\rho
=const=\rho _{0}=-1$\ leads this model to the Placzek one describing the
stimulated Raman scattering \cite{R20}. The typical magnitude of the
parameters are $\omega _{0}=10^{13}s^{-1}$, $M=10^{-22}g$, $\alpha
_{1}=10^{-15}cm^{2}$, $n_{A}|\rho _{0}|=10^{22}cm^{-3}.$

The steady state solutions of the Blomberge-Shen model (\ref{eq26}) are not
known. It was shown that under unidirectional approximation (when wave
propagates only in one of the possible directions) \cite{R21} this model can
be used to description of the self-induced transparency under two-photon
resonance.

\subsection{Isotropic solid dielectric model}

As pointed out in \cite{R22}, in the study of the ESP propagation in a solid
dielectric medium it is essential that the material equations adequately
account for dispersion of both the linear and the non-linear susceptibility
of the medium over practically its entire transparency range. Taking into
account the contributions of both the electron and phonon subsystems of the
solid dielectric one can describe normal and anomalous group dispersion.
Thus, the simplest generation of the Lorentz model was proposed 
\begin{equation}
\frac{\partial ^{2}P}{\partial t^{2}}+(\omega _{1}^{2}+\gamma
E^{2})P=n_{A}\alpha _{0}E+\beta RE,\quad \frac{\partial ^{2}R}{\partial t^{2}%
}+\omega _{2}^{2}R=\chi PE,  \label{eq27}
\end{equation}
where $P$ is electronic polarisation, the extra variable $R$ reflects
vibration motion of the atoms. Parameters $\alpha _{0},\beta ,\gamma ,\chi $
are the characteristic ones to phenomenological description of the
dielectric medium. The more complete model was proposed in \cite{R23}. The
generalised Lorentz model represented there by the set of equations 
\[
\frac{\partial ^{2}P_{e}}{\partial t^{2}}+\frac{2}{T_{e}}\frac{\partial P_{e}%
}{\partial t}+\omega _{e}^{2}P_{e}=\alpha _{e}E+\beta (R_{e}+R_{v})E, 
\]
\[
\frac{\partial ^{2}P_{i}}{\partial t^{2}}+\frac{2}{T_{i}}\frac{\partial P_{i}%
}{\partial t}+\omega _{i}^{2}P_{i}=\alpha _{i}E, 
\]
\begin{equation}
\frac{\partial ^{2}R_{e}}{\partial t^{2}}+\frac{2}{T_{e1}}\frac{\partial
R_{e}}{\partial t}+\omega _{e1}^{2}R_{e}=\gamma _{e}(P_{e}+P_{i})E,
\label{eq28}
\end{equation}
\[
\frac{\partial ^{2}R_{v}}{\partial t^{2}}+\frac{2}{T_{v}}\frac{\partial R_{v}%
}{\partial t}+\omega _{v}^{2}R_{v}=\gamma _{v}(P_{e}+P_{i})E, 
\]
where $P_{e}$\ and $P_{i}$ are the contributions to the polarisation of the
electronic and ionic components. The oscillators described by the variables $%
R_{e}$ and $R_{v}$ give rise to non-linear parametric coupling between the
ESP field and the medium. The dynamical parameter $R_{e}$ is responsible for
the electronic nonlinearity, and the dynamical parameter $R_{v}$ reflects
electron-phonon nonlinearity. The phenomenological parameters $\alpha _{e,i}$%
, $\omega _{e,i}$, $T_{e,i}$ characterise the dispersion of the electronic
and ionic linear polarisation responses. The coefficients $\beta $, $\gamma
_{e,v}$, $\omega _{e1,v}$, $T_{e1,v}$ characterise the dispersion of the
electronic and electron-vibrational non-linear responses, respectively.

A sequence of the approximations (it has been detailed in \cite{R23}) allows
to solve the equations (\ref{eq28}) that leads to non-linear wave equation 
\begin{equation}
\frac{\partial E}{\partial z}-a\frac{\partial ^{3}E}{\partial \tau ^{3}}%
+b\int\limits_{-\infty }^{\tau }Ed\tau ^{\prime }+gE^{2}\frac{\partial E}{%
\partial \tau }=\vartheta \Delta _{\perp }\int\limits_{-\infty }^{\tau
}Ed\tau ^{\prime },  \label{eq29}
\end{equation}
where $\tau =t-z/V$ is retarded time, $a$, $b$, $g$, and $\vartheta $\ are
positive parameters determined in \cite{R23}. $\Delta _{\perp }$ is the
transverse Laplacian. The Ref. \cite{R24} represents the vector generation
of the non-linear wave equation 
\begin{equation}
\frac{\partial \vec{E}}{\partial z}-a\frac{\partial ^{3}\vec{E}}{\partial
\tau ^{3}}+b\int\limits_{-\infty }^{\tau }\vec{E}d\tau ^{\prime }+g(\vec{E}%
\cdot \vec{E})\frac{\partial \vec{E}}{\partial \tau }+h\vec{E}\times (\vec{E}%
\times \frac{\partial \vec{E}}{\partial \tau })=0,  \label{eq30}
\end{equation}
describing propagation of the ESP of arbitrary polarisation. The numerical
simulations shown that the polarisation self-action of ESP consists in a
variation of the orientation of the vector of the electric field strength in
the direction to the propagation axis proportional to the field squared and
the velocity of rotation of its strength vector. There are no exact
analytical results, unfortunately.

\subsection{Gas of quantum particles}

A self-consistent model for a non-linear interaction of an intense extremely
short light pulse with a gas of atoms in framework of the quantum mechanics
has been discussed in \cite{R25}. The model includes the scalar wave
equation (\ref{eq11}) and one-dimensional Schr\"{o}dinger equation for an
electron, which is subjected to the intratomic potential $U(x)$ and the
optical radiation in the dipole approximation 
\[
i\hbar \frac{\partial \psi }{\partial t}+\frac{\hbar ^{2}}{2m}\frac{\partial
^{2}\psi }{\partial x^{2}}-U(x)\psi =exE\psi . 
\]
The polarisation $P$ of the medium, which consists of model one-electron
atoms with a number density $n_{A}$, is 
\[
P(t)=-en_{A}\int\limits_{-\infty }^{+\infty }x|\psi |^{2}dx. 
\]
By using the Schr\"{o}dinger equation we can rewrite the wave equation as 
\begin{equation}
\frac{\partial ^{2}E}{\partial z^{2}}-\frac{1}{c^{2}}\frac{\partial ^{2}E}{%
\partial t^{2}}=\frac{4\pi e^{2}n_{A}}{c^{2}m}E+\frac{4\pi en_{A}}{c^{2}m}%
\int\limits_{-\infty }^{+\infty }|\psi (x)|^{2}\frac{\partial U}{\partial x}%
dx.  \label{eqVanin}
\end{equation}
The first term describes the linear response of the free electrons in the
absence of the intratomic confining potential. The second term reflects the
contribution of the coupled electrons. In \cite{R25} it was shown by
numerical method that it is posible to generate a burst of harmonics with a
relative power (in the region of 1\%) and a duration amounting toonly a few
periods of the exciting optical field.

\subsection{Non-truncated Maxwell-Bloch model}

In order to investigate the self-induced transparency phenomena for
extremely short electromagnetic pulses in \cite{R26} \emph{the non-truncated
Maxwell-Bloch equations} have been proposed. These equations take the form 
\[
\frac{\partial ^{2}A}{\partial z^{2}}-\frac{1}{c^{2}}\frac{\partial ^{2}A}{%
\partial t^{2}}=-\frac{4\pi }{c}j-4\pi k^{4}\chi _{3}|A|^{2}A, 
\]
\begin{equation}
\frac{\partial j}{\partial t}-i(\omega _{0}+\gamma _{s}|A|^{2})j=i\frac{%
|m_{12}|^{2}}{\hbar c}\rho A,  \label{eq31}
\end{equation}
\[
\frac{\partial \rho }{\partial t}=\frac{2i}{\hbar c}\left( jA^{\ast
}-j^{\ast }A\right) , 
\]
where $A$ is a vector-potential of the electromagnetic wave, $j$ is atomic
transition current density, $\rho $\ is a population inverse density, $%
m_{12} $ is matrix element of resonant transition current. The Kerr and
Stark effects were taken into account by introduction of the terms
containing parameters $\chi _{3}$ and $\gamma _{s}$, respectively.

The results of the numerical simulations demonstrate the existence of the
self-similar solutions of the system of equations (\ref{eq31}). These
solutions should be called by the solitons because they are stable in
respect with pulse collisions. The main distinctive feature of the solitons
of the non-truncated Maxwell-Bloch equations in that they do not break-up
into the pulse sequence when the area of the incident pulse is higher then $%
2\pi $.

\section{Conclusion}

We have considered some case of model media that allow the explicit temporal
dependence of the electric field strength of the extremely short pulses to
be determined by analytically. We did not make any assumptions concerning
the harmonic carrier of the wave or the variation rate of the field in the
pulse. As usually, only steady state solutions describing the propagation of
the ESP can be found.

There are some investigations in the considered area that were left aside.
However, it seems useful to mention works \cite{R27, R28, R29} where the
propagation of a powerful electromagnetic pulses in a strongly non-linear
medium without dispersion has been considered. The exactly solvable models
are developed to describe the interaction of extremely short (few-cycle) 
\emph{transients} with certain classes of insulators and conductors.
Transient-excited fields are described analytically based on new, exact,
non-periodic and non-stationary solutions to Maxwell's equations, obtained
directly in time domain using a no Fourier-expansion, no time-space
separation method. Such non-separable solutions form the mathematical basis
of the non-periodic electromagnetic waves and make the standard harmonic
wave concepts of frequency, phase, refraction index, and phase velocity
irrelevant to the time description of non-periodic waves. Ref. \cite{R29}
reviews these investigations.

\section*{Acknowledgment}

We are grateful to Dr. S.V. Sazonov, Pr. S.A. Kozlov and Pr. A.V. Andreev
for valuable discussions. This work was supported by INTAS (European Union)
under the grant No. 96-0339.

\end{document}